\documentclass[12pt]{iopart}
\usepackage{amssymb}
\usepackage{iopams}
\usepackage{graphicx}
\usepackage{color}
\usepackage{ulem} 
\newcommand{\lacatbcodve}{La$_{0.6}$Tb$_{0.2}$Ca$_{0.2}$CoO$_3$}
\newcommand{\lacatbcox}{La$_{\rm 0.8-x}$Tb$_{\rm x}$Ca$_{0.2}$CoO$_3$}
\newcommand{\lasrcox}{La$_{\rm 1-x}$Sr$_{\rm x}$CoO$_3$}
\newcommand{\lacacox}{La$_{\rm 1-x}$Ca$_{\rm x}$CoO$_3$}
\newcommand{\labacox}{La$_{\rm 1-x}$Ba$_{\rm x}$CoO$_3$}

\newcommand{\laco}{LaCoO$_3$}

\newcommand{\rtric}{$R\overline{3}c$}
\newcommand{\cok}{CoO$_6$}
\newcommand{\st}{$\rm ^o$}
\newcommand{\stc}{$\rm ^oC$}
\newcommand{\teLSiii}{$t_{2g}^6e_g^0$}

\newcommand{\teISiii}{$t_{2g}^5e_g^1$}

\newcommand{\teHSiii}{$t_{2g}^4e_g^2$}
\newcommand{\figwl}{0.75} 
\newcommand{\figwp}{0.75} 
\newcommand{\figws}{0.85} 
\newcommand{\tabws}{0.95} 

\begin{document}
\sloppy
\title[Effect of Ising-type Tb$^{3+}$ on the magnetism of La,Ca cobaltite]{
Effect of Ising-type Tb$^{3+}$ ions on the low-temperature magnetism of La, Ca cobaltite.}
\author{
 K. Kn\'{\i}\v{z}ek$^1$, Z. Jir\'{a}k$^1$, J. Hejtm\'{a}nek$^1$, M. Veverka$^1$
 O. Kaman$^1$, M. Mary\v{s}ko$^1$, E. \v{S}antav\'{a}$^1$,
 G. Andr\'{e}$^2$}
\address{
 $^1$Institute of Physics ASCR, Cukrovarnick\'a 10, 162 00 Prague 6, Czech Republic. \\
 $^2$Laboratoire L\'{e}on Brillouin, CEA-CNRS, CEA-Saclay, 91191 Gif-s-Yvette Cedex, France.
}
\begin{abstract}
Crystal and magnetic structures of the $x=0.2$ member of La$_{\rm 0.8-x}$Tb$_{\rm
x}$Ca$_{0.2}$CoO$_3$  perovskite series have been determined from the powder neutron
diffraction. Enhancement of the diffraction peaks due to ferromagnetic or cluster glass ordering
is observed below $T_C=55$~K. The moments evolve at first on Co sites, and ordering of
Ising-type Tb$^{3+}$ moments is induced at lower temperatures by a molecular field due to Co
ions. The final magnetic configuration is collinear F$_x$ for cobalt subsystem, while it is
canted F$_x$C$_y$ for terbium ions. The rare-earth moments align along local Ising axes within
\textit{ab}-plane of the orthorhombic $Pbnm$ structure. The behavior in external fields up to
$70-90$~kOe has been probed by the magnetization and heat capacity measurements. The dilute
terbium ions contribute to significant coercivity and remanence that both steeply increase with
decreasing temperature. A remarkable manifestation of the Tb$^{3+}$ Ising character is the
observation of a low-temperature region of anomalously large linear term of heat capacity and
its field dependence. Similar behaviours are detected also for other terbium dopings $x=0.1$ and
0.3.
\end{abstract}
%
\maketitle

\section{Introduction}

The properties of perovskite cobaltites are largely affected by closeness in energy of different
local states of the octahedrally coordinated cobalt ions. The well-known example is \laco, in
which a spin transition, or spin-state crossover, occurs in the course of temperature. It has been
recognized that the \laco\ ground state is based on non-magnetic low spin state of Co$^{3+}$ (LS,
$S=0$, \teLSiii), while with increasing temperature above $\sim 40$~K, the energetically close
high spin Co$^{3+}$ (HS, $S=2$, \teHSiii) species start to be populated by thermal excitation
\cite{RefGoodenough1958JPCS6_287}. The HS population is practically saturated above 150~K, making
about 40-50\%. Strong HS/LS nearest neighbor correlations or even short-range ordering are
anticipated in that phase
\cite{RefBari1972PRB5_4466,RefKyomen2005PRB71_024418,RefKnizek2009PRB79_014430GGA,RefKrapek2012PRB86_195104}.
At still higher temperature the ordering melts. This process is accompanied with a drop of
electrical resistivity centered at about 530~K, which can be interpreted as a transition from Mott
insulator to a quasi-metallic state. The high-temperature phase of \laco\ should be regarded as
inhomogeneous, since it essentially retains the LS/HS Co$^{3+}$ disproportionation with only short
visits to intermediate spin (IS, $S=1$, \teISiii) configurations \cite{RefKrapek2012PRB86_195104}.

While the pure \laco\ is paramagnetic at all temperatures, the doping of holes by chemical
substitution brings generally ferromagnetic interactions and leads finally to the long-range
ferromagnetic (FM) order. This evolution is best documented for the \lasrcox\ system
\cite{RefHe2009EPL87_027006,RefHe2009PRB80_214411}. The mildly doped compounds exhibit a
non-uniform ground state with hole-rich FM regions of IS/LS character for Co$^{3+}$/Co$^{4+}$,
embedded in the hole-poor matrix with main weight of LS Co$^{3+}$. The origin of such two-phase
competition cannot be understood within any standard band picture, and is intimately associated
with strongly correlated nature of the materials
\cite{RefSboychakov2009PRB80_024423,RefSuzuki2009PRB80_054410}. The transition to a more
homogeneous state is observed at  $x=0.22$ as documented by onset of metallic conductivity and
finite electronic heat with linear coefficient $\gamma\sim 40$~mJ/K$^{2}$
\cite{RefMuta2002JPSJ71_2784}. Above this critical concentration, the \lasrcox\ compounds show
characteristics of conventional ferromagnets, namely the large $\lambda$-anomaly in the specific
heat at $T_C$ and the critical behavior manifested by a sharp peak in small-angle neutron
scattering and by critical exponents $\beta$, $\gamma$ and $\delta$ belonging to universality
class of 3D Heisenberg model
\cite{RefHe2009EPL87_027006,RefHe2009PRB80_214411,RefKhan2010PRB82_064422}.

The situation in calcium doped system \lacacox\ is less explored. In this case a much more uniform
FM phase seems to form starting from weak doping, though the saturated moments are low and no
metallic conductivity is reached
\cite{RefTaguchi1982JSSC41_329,RefMuta2002JPSJ71_2784,RefKriener2004PRB69_094417}. There are,
nevertheless, few reports pointing to a certain phase separation, similarly to what is known for
\lasrcox. In particular, it is demonstrated on \lacacox\ single crystals ($x=0.1-0.2$) that their
magnetic state evolves from ensemble of weakly interacting spin clusters to typical cluster glass
\cite{RefSzymczak2005JMMM285_386}.  The ferromagnetic phase is formed at higher Ca content, in
particular below 170~K for $x=0.3$ \cite{RefMuta2002JPSJ71_2784}. Interestingly, a reentrant spin
glass transition has been observed at 100~K, and magnetic relaxation experiments have shown that
both the ferromagnetic and the reentrant spin-glass phases in $x=0.3$ are nonequilibrium states,
which exhibit a magnetic aging characteristic for spin glasses or disordered and frustrated
ferromagnets \cite{RefKundu2005PRB72_144423}.

In order to understand better the behavior of \laco-related compounds with the calcium doping, we
have undertaken the study of the \lacatbcox\ system. The Tb$^{3+}$ ions have been selected since
they show complex magnetic behavior due to crystal field effects, and can be effectively used as
local probe of the low-temperature magnetism produced by cobalt subsystem. We have found, with the
help of the neutron diffraction, magnetic measurements and investigation of low-temperature heat
capacity, that the magnetic ground state is highly non-uniform and Ising-type Tb$^{3+}$ moments
influence critically the magnetic properties.

\section{Experimental}

Samples \lacatbcox\ ($x=0.1$, 0.2 and 0.3) were synthesized by a sol-gel procedure followed by
annealing at 1000\stc\ under air for at least 24 hours. The starting materials included
stoichiometric amounts of La$_2$O$_3$, Tb$_4$O$_7$, CaCO$_3$ and Co(NO$_3$)$_2$ solutions with
chemically determined metal content while ethylene glycol was used to prepare the gel precursor.
The final products were checked for phase purity by X-ray diffraction. The perovskite crystal
structures are of the orthorhombic $Pbnm$ type, except for a small coexistence of the \rtric\
phase in the $x=0.1$ sample. More detailed structural investigation was performed by neutron
diffraction in LLB (Saclay, France) on the G41 diffractometer using a wavelength of 2.422~\AA. The
neutron diffraction data, obtained at selected temperatures down to 1.8~K, were analyzed by a
Rietveld method with the help of the FULLPROF program (version 4.80/2010).

For a complex physical characterization the \mbox{$x=0.2$} compound (\lacatbcodve) was selected.
The magnetic moment was measured on a SQUID magnetometer (MPMS-XL; Quantum Design) over the range
$2-400$~K under various field 100~Oe~-~70~kOe. The zero-field-cooled (ZFC) and field-cooled (FC)
runs were performed. The hysteresis loops were measured up to maximum field of 70~kOe at selected
temperatures starting from $T=2$~K upwards. The remanent magnetization $vs.$~$T$ dependence was
measured after recording the hysteresis loop $T=2$~K and decreasing $H$ to zero. The ZFC long-time
relaxation experiments were performed by cooling the sample from $T=300$~K to a target temperature
$T$ and keeping it for a certain wait time $t_w$, typically $10^3$~sec, before setting the field
$H=10$~Oe and starting the measurement. The time evolution of the magnetic moment was then
recorded between $t=0$ and $t=4900$~sec. The AC susceptibility measurements were carried out with
the amplitude of driving field 3.9~Oe and frequency $0.3-300$~Hz.

The specific heat was measured by PPMS device (Quantum Design) using the two$-\tau$ model. The
data at zero field and under fields up to 90~kOe were collected generally on sample cooling. The
experiments at very low temperatures (down to 0.4~K) were done using the He$^3$ option.

\section{Results and discussion}

\subsection{Transport properties}

\begin{figure}
\includegraphics[width=\figwp\columnwidth,viewport=0 0 582 819,clip]{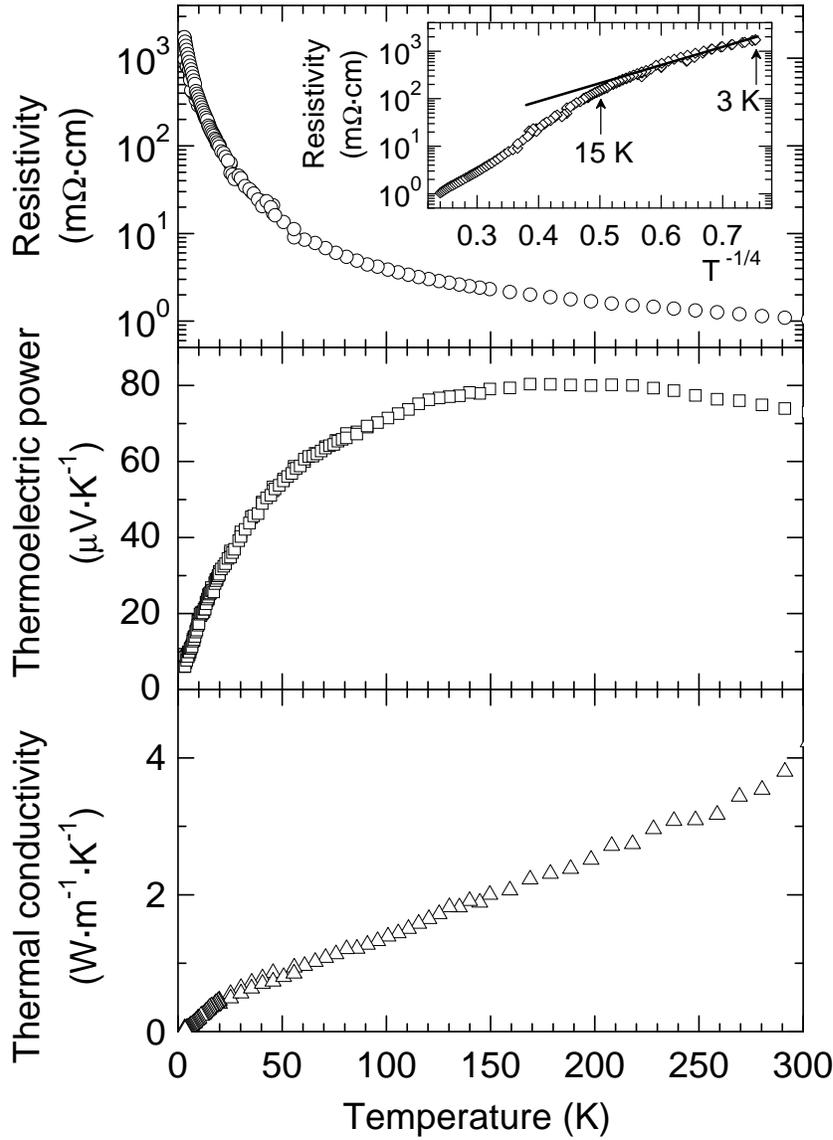}
\caption{Resistivity, thermoelectric power and thermal conductivity of \lacatbcodve. The linear
part of resistivity $vs.$ T$^{-1/4}$ dependence, seen in the inset of upper panel for $T<15$~K, is
characteristic for the variable range hopping.} \label{FigTransport}
\end{figure}

As a first physical characterization of the \lacatbcodve\ system, the basic transport phenomena
were probed. Fig.~\ref{FigTransport} presents the temperature dependence of electrical
resistivity, thermoelectric power and thermal conductivity. The displayed data show that the
electronic transport is of a transitional behavior between the strongly activated semiconductor
\laco\ and the doped ferromagnetic metallic compounds \lasrcox. The actual regime, manifested by
the activated resistivity and linear (metallic-like) thermopower at the lowest temperatures, can
be described as the variable range hopping. This type of behavior is typical for systems with
localized electronic states forming a quasi-continuous band around Fermi level;  for doped
cobaltites see \textit{e.g.} \cite{RefHejtmanek2010PRB82_165107}. The resistivity above $\sim15$~K
follows a non-standard power law dependence, similarly to what was reported earlier for some
polycrystalline specimens of electron-doped LaCoO$_3$ \cite{RefJirak2008PRB78_014432} or mixed
systems LaCo$_{\rm 1-x}$Cu$_{\rm x}$O$_3$ \cite{RefDlouha2011JSNM24_741}.

\subsection{Crystal structure}

\begin{figure}
\includegraphics[width=\figwp\columnwidth,viewport=0 180 582 819,clip]{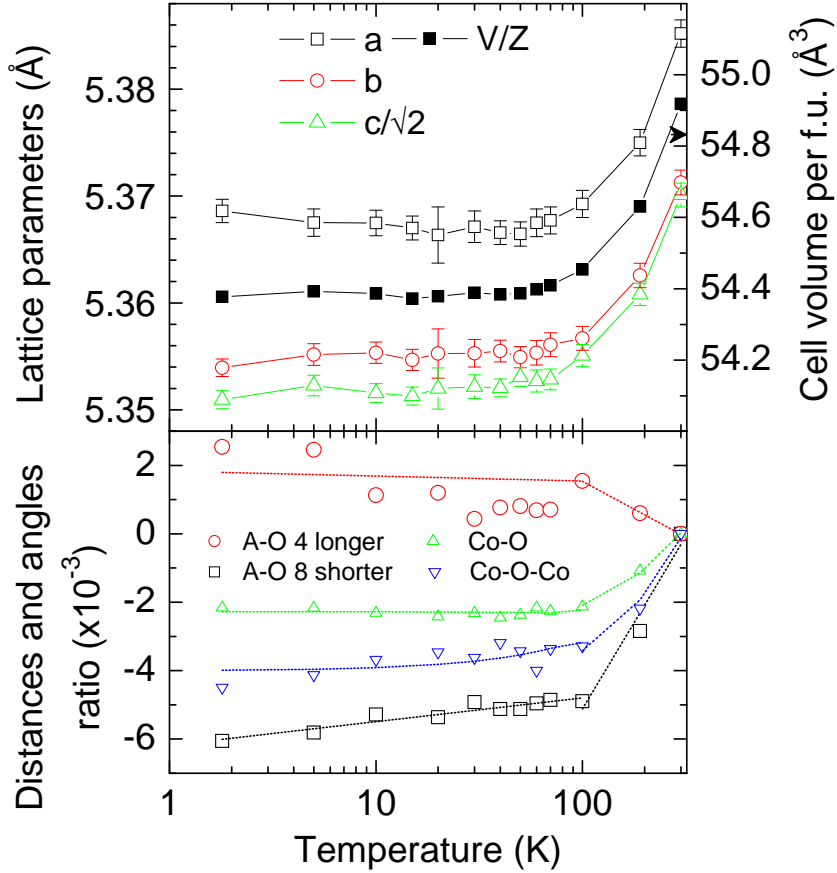}
\caption{(a) Lattice parameters and cell volume/f.u. dependence on temperature. (b) The relative
changes of short and long A-O bonds, A=La$_{0.6}$Tb$_{0.2}$Ca$_{0.2}$, together with average
values of Co-O bonds and O-Co-O angles.} \label{FigNDLatt}
\end{figure}

\begin{table}
\caption{The crystallographic data summary for \lacatbcodve\ at 1.8, 100 and 300~K. Space group
$Pbnm$. Atom coordinates: LaTbCa $4c$(x,y,1/4), Co $4b$(1/2,0,0), O1 $4c$(x,y,1/4), O2
$8d$(x,y,z).}

\begin{tabular*}{\tabws\columnwidth}{@{\extracolsep{\fill}} l|rrr}
\hline
T (K)    & 1.8        & 100        & 300        \\
\hline
a (\AA)  & 5.3686(5)  & 5.3693(6)  & 5.3852(6)  \\
b (\AA)  & 5.3539(4)  & 5.3567(6)  & 5.3713(6)  \\
c (\AA)  & 7.5674(6)  & 7.5732(7)  & 7.5945(8)  \\
\hline
x,LaTbCa & -0.0005(11) & -0.0075(8) & -0.0077(8) \\
y,LaTbCa & 0.0294(5)   & 0.0279(4)  & 0.0262(5)  \\
x,O1     & 0.0750(13)  & 0.0634(9)  & 0.0618(9)  \\
y,O1     & 0.4911(7)   & 0.4912(6)  & 0.4930(7)  \\
x,O2     & -0.2775(8)  & -0.2793(7) & -0.2803(7) \\
y,O2     & 0.2813(7)   & 0.2793(6)  & 0.2770(8)  \\
z,O2     & 0.0303(6)   & 0.0345(4)  & 0.0339(4)  \\
\hline
\end{tabular*}

\label{TabStruktura}
\end{table}

The crystal and magnetic structure of \lacatbcodve\ perovskite was determined from the powder
neutron diffraction, measured at the temperatures from 1.8 to 300~K. Let us note that, without
terbium substitution, the crystal structure of \lacacox\ changes from rhombohedral (space group
\rtric) for $x<0.1-0.2$ to orthorhombic (space group $Pbnm$) for $x>0.2-0.3$, with both phases
coexisting in the intermediate doping range
\cite{RefBurley2004PRB69_054401,RefMastin2006CHEMMAT18_1680}. For \lacatbcodve\ the single
orthorhombic $Pbnm$ structure is observed over the whole experimental temperature range. For three
selected temperatures the complete data including atomic coordinates are presented in
Table~\ref{TabStruktura}. The occupancy of oxygen sites is refined to 1.009$\pm0.003$, pointing to
oxygen content in the \lacatbcodve\ sample slightly above the ideal stoichiometry. This result may
suggest that the cobalt valence in \lacatbcodve\ is shifted from the formal mixture
0.2~Co$^{4+}$/0.8~Co$^{3+}$ to actual 0.25~Co$^{4+}$/0.75~Co$^{3+}$. The goodness of Rietveld fit
is characterized by values
 $R_{p}=2.03$\%, $R_{wp}=2.77$\% and $R_{Bragg}=1.92$\% at 300~K, and
 $R_{p}=2.85$\%, $R_{wp}=4.01$\%,    $R_{Bragg}=2.06$\% and $R_{mag}=9.17$\% at 1.8~K.

The temperature evolution of lattice parameters and cell volume \textit{per} f.u. is displayed in
the upper panel of Fig.~\ref{FigNDLatt}. The relation between lattice parameters, $a>b \sim c/
\sqrt 2$,  is different from the situation in perovskites ABO$_3$ with smaller A cations (and
smaller tolerance factor), for which the tilting of the octahedra network is the dominant source
of the orthorhombic distortion, and the relation between lattice parameters $b > c/\sqrt 2 > a$ is
typical. The present type of orthorhombic deformation should be thus related to a small sheer
distortion of the \cok\ octahedron, namely the deviation of O-Co-O angles from 90\st, that
prevails over effects of the octahedral tilting.

The extent of octahedral tilting is quantified by the average bond angle Co-O-Co, which makes
160\st\ at the room temperature, decreases to 159.5\st\ at 100~K and then to 159.2\st\ at the
lowest temperature. The average Co-O bond distance decreases from 1.930~\AA\ at 300~K to
1.926~\AA\ at 100~K, but then it is approximately constant down to the lowest temperature. The
\cok\ octahedron is almost regular, the difference between the longest and shortest Co-O distance
does not exceed 0.02~\AA. The three non-equivalent O-Co-O angles in $Pbnm$ structure range between
$88.8-89.6$\st.

The change of \cok\ tilting incurred by lowering temperature is also reflected in the temperature
evolution of coordination sphere of the big cation (La$_{0.6}$Tb$_{0.2}$Ca$_{0.2}$) in the A site
of perovskite structure. The 12 bonds could be divided into two groups of 8 shorter and 4 longer
distances, having different evolution with temperature, see the lower panel of
Fig.~\ref{FigNDLatt}. The 8 shorter bonds shorten relative to room temperature values, whereas the
4 longer bonds elongate compared to room temperature. The temperature evolution of the shorter A-O
bonds resembles closely that of Co-O-Co bond angle.

\subsection{Magnetic properties}

\begin{figure}
\includegraphics[width=\figws\columnwidth,viewport=0 250 582 819,clip]{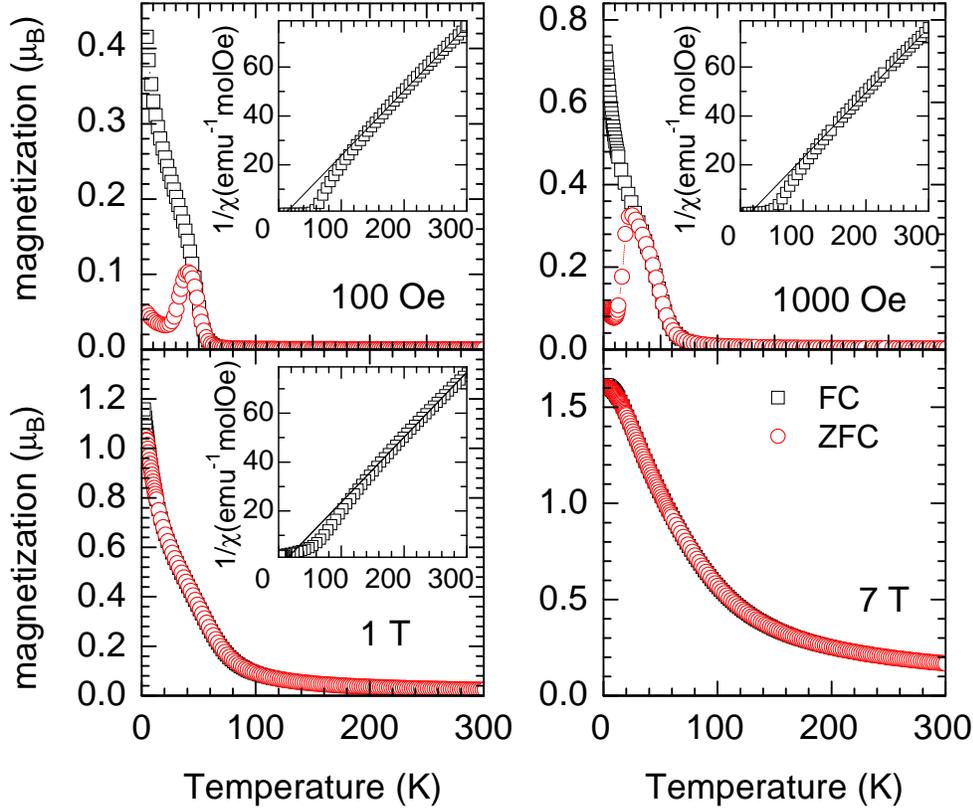}
\caption{The magnetic moment of \lacatbcodve\ measured under various fields
in the FC ($\square$) and ZFC(\textcolor{red}{$\bigcirc$}) regimes.
The insets show the corresponding inverse susceptibility graphs.} \label{FigZFCFC}
\end{figure}

Magnetic properties of \lacatbcodve\ were probed by the DC magnetization measurements and by the
frequency dependent AC susceptibility experiments. The temperature curves of magnetic moment
measured under various fields from 100~Oe to 70~kOe are presented in Fig.~\ref{FigZFCFC}. The
moments observed above $\sim 60$~K are proportional to applied field, which is a signature of
paramagnetic state. The paramagnetic properties are more apparent in the insets of
Fig.~\ref{FigZFCFC} where data are plotted in terms of inverse susceptibility. With increasing
temperature the Curie-Weiss dependence $\chi=C/(T-\theta)$ is approaching, where the molar value
of Curie constant depends on the effective paramagnetic moment, $C=N\mu_{eff}^2/3k$. The fit
yields $\mu_{eff}^2 = 28.6$~$\mu_B^2$ per f.u. and Weiss temperature $\theta = 23$~K. The square
value of the effective moment agrees with a theoretical expectation, namely the weighted sum of
$\mu_{eff}^2$ for 0.2 Tb$^{3+}$ (the free-ion value $\mu_{eff}=9.72$~$\mu_B$) and valence mixture
0.25~Co$^{4+}$/0.75~Co$^{3+}$, both in the intermediate spin state ($S=1.5$ and 1, respectively),
which gives total $\mu_{eff}^2 = 28.7$~$\mu_B^2$. The positive Weiss $\theta$ is attributed to FM
interaction between Co ions, since the interaction between more diluted Tb ions is supposedly weak
and of antiferromagnetic (AFM) type.

\begin{figure}
\includegraphics[width=\figwl\columnwidth,viewport=0 400 582 819,clip]{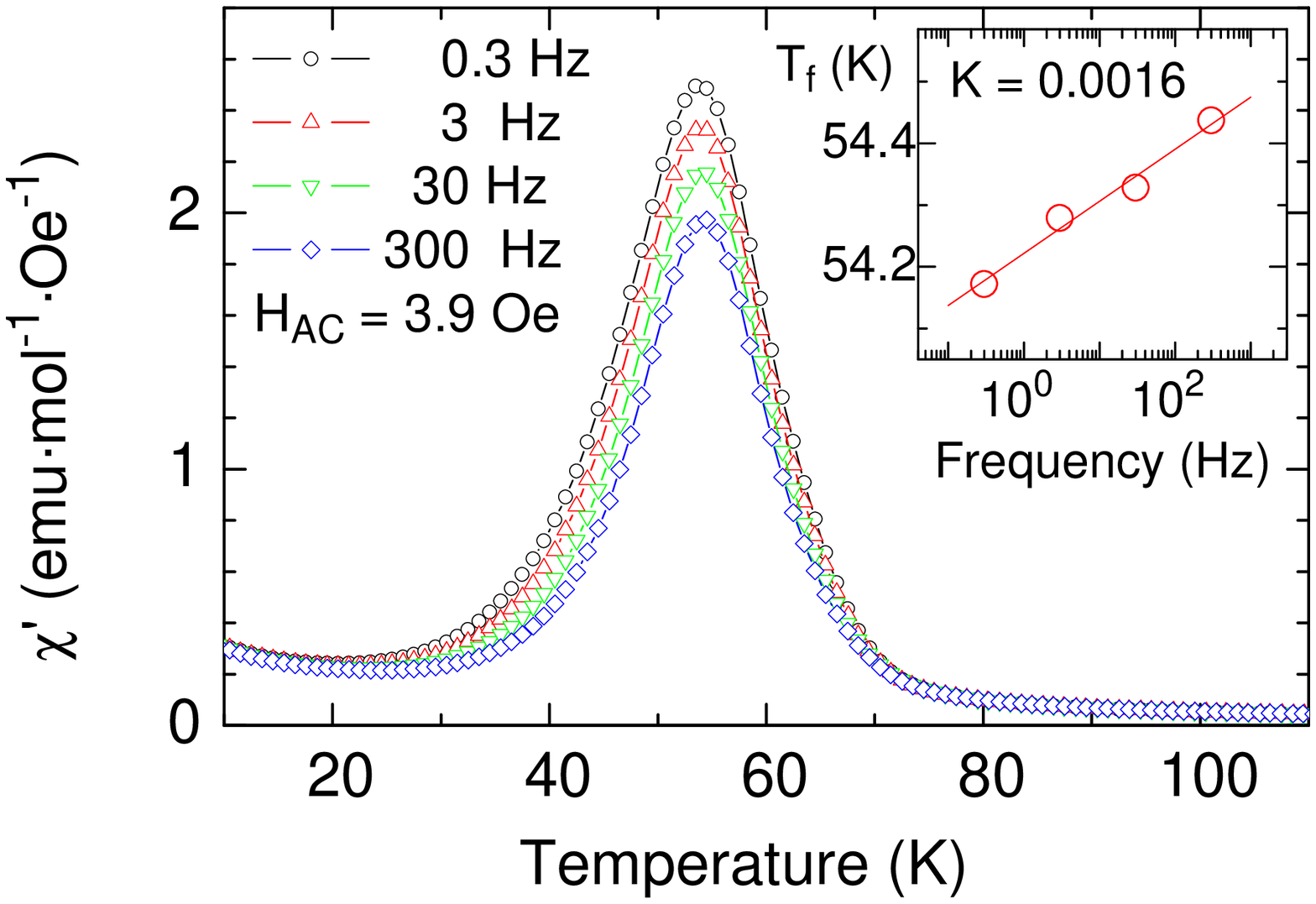}
\caption{Real part of AC susceptibility in \lacatbcodve.} \label{FigACreal}
\end{figure}

Below $\sim 60$~K the magnetic moment rapidly increases. The low-temperature diverging tail of FC
curves, clearly observed in Fig.~\ref{FigZFCFC} up to 10~kOe, can be assigned mainly to the
magnetic response of Tb$^{3+}$ ions. There is an inflection point at about 20~K, which suggests
that this paramagnetic-like term is superimposed on another term due to Co$^{3+}$/Co$^{4+}$ ions,
having a form close to Brillouin curve. This latter contribution and its gradual saturation in
high fields remind the behaviour of a conventional ferromagnet. However, the bifurcation between
ZFC and FC curves, which shifts with increasing field to lower temperatures, indicates rather the
formation in \lacatbcodve\ of a non-uniform state of glassy character. This conclusion is further
supported by AC susceptibility measurements over the frequency range $0.3-300$~Hz, presented in
Fig.~\ref{FigACreal}. As shown in the inset, the characteristic temperature $T_f$, at which the
real part $\chi '$ passes through a maximum, exhibits with increasing frequency $\nu$ of the
applied AC field an upward shift that can be quantified by a semiempirical dimensionless parameter
$K = \Delta T_f/[T_f\Delta (log\nu)]= 0.0016$. Such weak frequency dependence, compared to much
larger $K$-values in canonical spin-glass systems \cite{RefMydosh1993}, suggests that FM clusters
that freeze near $T_f\sim55$~K are relatively large.

Typical manifestation of the frustrated magnetic state is the long-time relaxation of
magnetization and ageing effects. The data we obtained are nearly identical with results of
Nam~\etal\ on polycrystalline La$_{0.5}$Sr$_{0.5}$CoO$_3$ exhibiting similar non-uniform magnetic
state in which the FM and glassy behaviors coexist \cite{RefNam1999PRB59_4189}. In particular, for
$T=40$~K, \textit{i.e.} slightly below the freezing temperature, the relaxation of moment after
application of probing field $H=10$~Oe could be described by the stretched exponential function
and the maximum of derivative of $m(T)$ with respect to $\ln(T)$ is attained at the elapsed time
close to the wait time $t=t_w$ ($t_w=1000$, 2000 and 3000~sec).

The magnetization curves for $T\leq40$~K are presented in Fig.~\ref{FigVirgin}. In the temperature
range down to $\sim 15$ K the magnetization shows a rapid rise at lower fields, but the next
increase is more gradual and lacks saturation in the highest field of 70~kOe. All these features
differ from the conventional behavior of bulk FM and are typical for cluster glass with broad size
distribution of FM domains and strong AFM interactions between them. Below $\sim 15$ K, an
anomalous behavior is observed. The virgin magnetization curves are specified by a linear initial
part followed by a break after which a steep rise, resembling a metamagnetic transition, occurs.
The critical field corresponding to the break increases with decreasing temperature in a
hyperbolic way as seen in the inset of Fig.~\ref{FigVirgin}. Since our neutron diffraction study
does not indicate any change of magnetic ground state below 15~K, the existence of the
metamagnetic transition can be discarded. Instead, we relate the observed character of
magnetization curves to the low-temperature effect of Tb$^{3+}$ moments. Namely, the local
anisotropy connected with the presence of Tb$^{3+}$ ions may lead to the pinning of domain walls
in FM ordered regions, which influences both the coercivity seen in hysteresis loops
(Fig.~\ref{FigHysteresis}) and the metamagnetic-like form of virgin curves. Let us note that the
coercive field reaches $\sim 7$~kOe at 2~K, and the above mentioned "metamagnetic transition" is
observed at the same field.

We may conclude this section by a statement that the chemically highly inhomogeneous compound
\lacatbcodve\ shows a spectrum of behaviors, some of which are typical for bulk FM and others
remind glassy systems. Apart of clear manifestations of long-range order like open hysteresis
loops or survival of a finite remanent magnetization up to 55~K (see the inset of
Fig.~\ref{FigHysteresis}), there is an evidence of short- and long-time relaxation processes.

\subsection{Magnetic ordering}

\begin{figure}
\includegraphics[width=\figwl\columnwidth,viewport=0 400 582 819,clip]{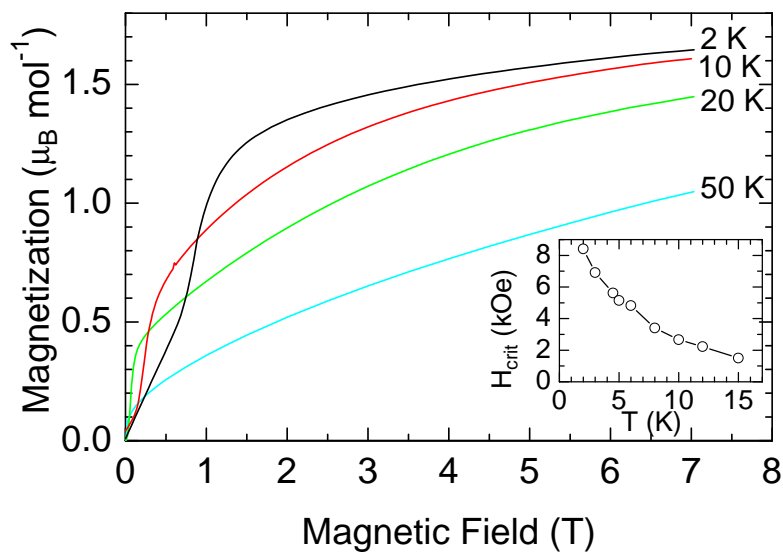}
\caption{Virgin magnetization curves of \lacatbcodve. The inset shows the temperature evolution
of critical field of the "metamagnetic transition".} \label{FigVirgin}
\end{figure}
\begin{figure}
\includegraphics[width=\figwl\columnwidth,viewport=0 400 582 819,clip]{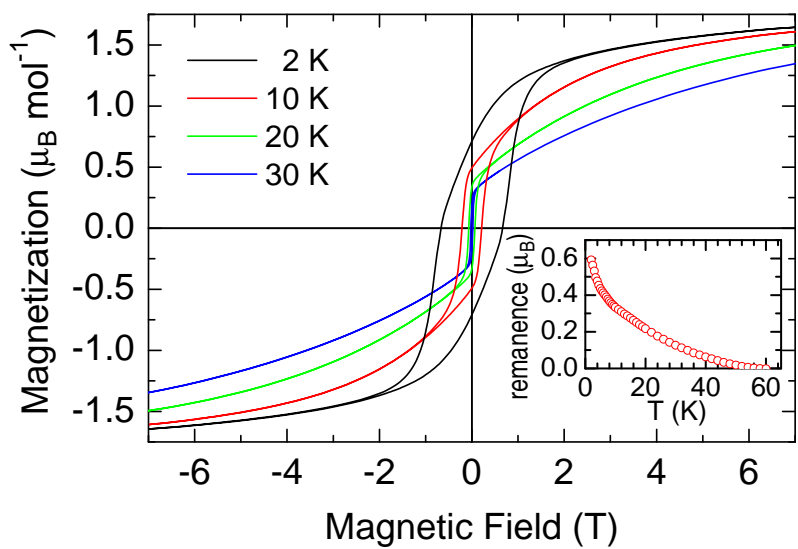}
\caption{Hysteresis magnetization loops of \lacatbcodve. The inset shows the temperature dependence
of remanent magnetization.} \label{FigHysteresis}
\end{figure}

\begin{figure}
\includegraphics[width=\figwl\columnwidth,viewport=0 200 582 819,clip]{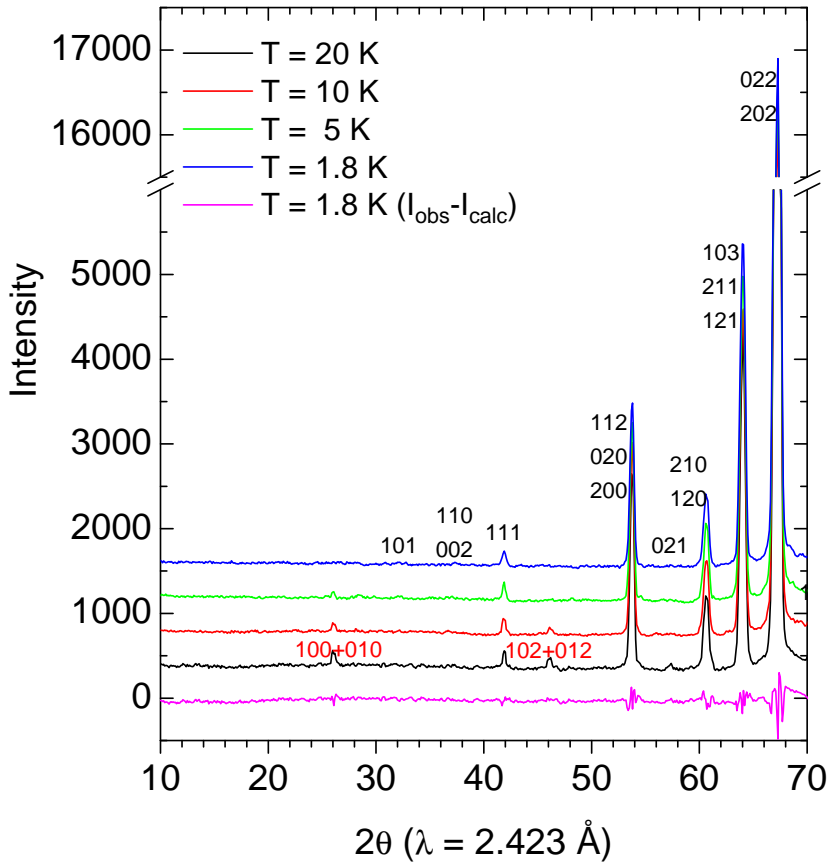}
\caption{The low-angle part of neutron diffraction patterns taken on \lacatbcodve\
at temperatures $1.8-20$~K. The FM ordering is deduced from
hardly visible enhancement of 110+002 and 112+020+200 lines, the C-type AFM order is manifested by
the additional magnetic reflections, 100+010 and 102+012.} \label{FigND}
\end{figure}

The neutron diffraction data of \lacatbcodve\ at selected low temperatures are displayed in
Fig.~\ref{FigND}. Presence of magnetic ordering is generally manifested in the neutron diffraction
data as an enhancement of some low-angle diffraction peaks. In closer inspection, one may indeed
notice a very weak peak at the calculated position of (110+002) reflection, whose nuclear
contribution is accidentally zero for present compound. This peak emerges below $\sim50$~K and, at
the same temperature, a small enhancement of the relatively strong 112+020+200 reflection can be
detected. Such additional intensity is an indication of magnetic contribution due to long-range FM
ordering of cobalt spins at least in a part of the sample. A further change can be observed below
20~K. First, the intensity of 112+020+200 starts to increase more rapidly, while that of 110+002
practically vanishes. The opposite changes on these peaks suggest that also the terbium moments at
A-sites adopt FM ordering, and their orientation is parallel to the cobalt ones. Namely, the
intensity of neutron diffraction peak 110+002 drops since it is given by square of
$m_{Co}-m_{Tb}$, while the intensity 112+020+200 given by square of $m_{Co}+m_{Tb}$ raises.

The second, well marked change below 20~K is the appearance of new magnetic peaks at 100+010 and
102+012 positions. These peaks are indicative for the C-type AFM arrangement in the sample and may
originate either in magnetic phase coexistence or in spin canting. Although a partial AFM order of
cobalt spins was already detected by neutron diffraction for some cobaltites of similar doping,
like \labacox\ ($x=0.17-0.19$) \cite{RefTong2011PRL106_156407}, we relate the present observation
to the canted arrangement of Tb$^{3+}$ moments within the FM phase. Let us note that the canting
is an inevitable consequence of Ising character of the Tb$^{3+}$ moments and alternation of easy
axes in the \textit{ab}-plane of the orthoperovskite structure \cite{RefGruber2008JLUM128_1271}.
More known example is the TbCoO$_{3}$ cobaltite with non-magnetic LS Co$^{3+}$, which shows below
T$_N\sim3.5$~K a spontaneous Tb$^{3+}$ ordering also of canted type, $A_xG_y$ in Bertaut's
notation \cite{RefKappatsch1970JPHYSFRAN31_369,RefMunoz2012EJICH2012_5825}.

\begin{figure}
\includegraphics[width=\figwl\columnwidth,viewport=0 400 582 819,clip]{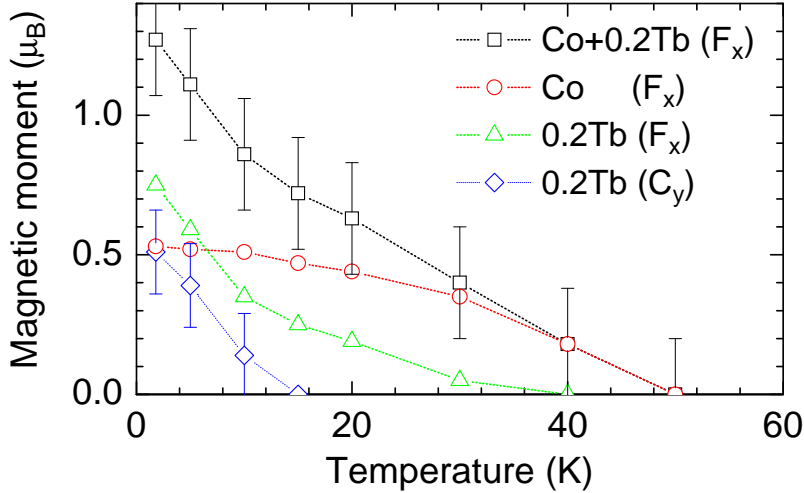}
\caption{FM ordered magnetic moments of Co and Tb and the C-type AFM ordered magnetic moments of Tb.
Error bars for the partial FM components are similar to those shown for
the overall Co+0.2Tb moment, $\sim\pm 0.2\mu_B$.}
\label{FigCoTb}
\end{figure}

The results of the Rietveld refinement are presented in Fig.~\ref{FigCoTb}. The spontaneous
ordering is established  at first on Co sites, and the ordering of the Tb moments arises gradually
at lower temperatures by action of molecular field induced by Co ions. The crystallographic
orientation of Co moments cannot be unambiguously determined due to small splitting of reflections
of the pseudocubic perovskite cell. We suppose that the magnetic axes of FM domains are
distributed rather randomly, but below 20~K the Co spins align in accordance with the easy local
axes of Ising-type Tb$^{3+}$ ions, which are within the \textit{ab}-plane, inclined to $\pm36$\st\
out of \textit{b}-direction. Based on this angle, found actually on related perovskite system
TbAlO$_3$ \cite{RefGruber2008JLUM128_1271}, and using the observed moments at 1.8~K in
Fig.~\ref{FigCoTb}, we may conclude that the final configuration is of collinear $F_x$ type with
long-range ordered moment of $0.55\mu_B$/Co, and the rare-earth ordered moments form the canted
arrangement F$_x$C$_y$ with average components \textit{per} site $m_F=0.75\mu_B$ and
$m_C=0.5\mu_B$, or \textit{per} rare-earth ion, $m_F=3.8\mu_B$/Tb and $m_C=2.5\mu_B$/Tb. This
means that the terbium moments are inclined from the $a$-axis to 34\st\ and their ordered
magnitudes make $4.5\mu_B$. The orientation of canted arrangement with respect to the
crystallographic axes then tells that the AFM diffraction lines observed at 100+010 and 102+012
positions in the $T=1.8$~K diffraction pattern in Fig.~\ref{FigND} are in fact the (100) and (102)
intensities, while (010) and (012) are extinguished. As the bulk magnetic moment is concerned, the
sum of FM components of the Co and Tb ions, $1.3~\mu_B$ \textit{per} f.u. at 1.8~K, seems to
correspond very well to the magnetization data in Figs.~\ref{FigVirgin} and \ref{FigHysteresis}
when an intuitive extrapolation to zero field is done.

\begin{figure}
\includegraphics[width=\figwl\columnwidth,viewport=0 200 582 819,clip]{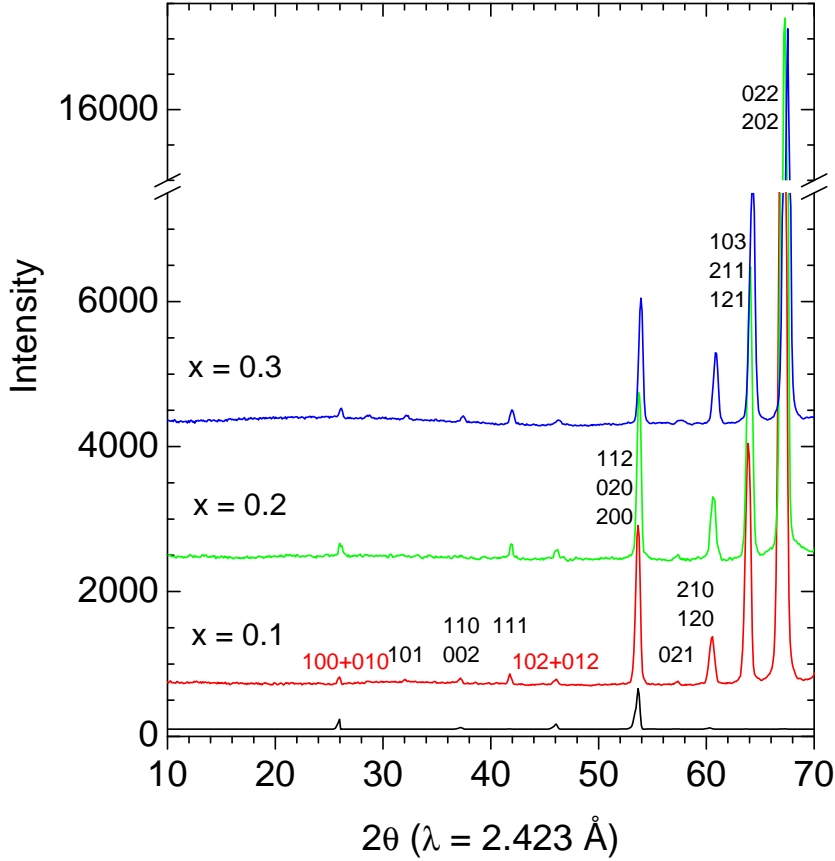}
\caption{Comparison of neutron diffraction patterns taken on
the \lacatbcox\ series ($x=0.1$, 0.2 and 0.3) at 1.8~K.
For $x=0.1$, the lower curve displays calculated magnetic intensities,
namely the AFM lines 100+010 and 102+012 and
FM contribution to structural lines 110+002 and 112+020+200.}
\label{FigNDall}
\end{figure}

The same set of FM and AFM reflections and, consequently, very similar magnetic structures (see
Fig.~\ref{FigStruktura} below) have been observed also for other terbium dopings in our
\lacatbcox\ series. For a comparison, the low-angle parts of neutron diffraction patterns of
$x=0.1$, 0.2 and 0.3 at 1.8~K are presented together in Fig.~\ref{FigNDall}. The refined moments
at cobalt sites and the magnitudes of terbium moments and their inclination from the orthorhombic
axis $a$ are summarized in Table~\ref{TabMag}.

\begin{table}
\caption{Data summary on magnetic ordering in the \lacatbcox\ series.}
\begin{tabular*}{\tabws\columnwidth}{@{\extracolsep{\fill}} l|rrr}
\hline
            & $x=0.1$    & $x=0.2$    & $x=0.3$    \\
\hline
T$_C$(K)    & 80         & 55         & 30         \\
\hline
$\mu_B$/Co      & 1.05 & 0.55 & 0.35 \\
$\mu_B$/Tb      & 8.4  & 4.5  & 3.0  \\
$\varphi$(\st)  & 33   & 34   & 31   \\
\hline
\end{tabular*}
\label{TabMag}
\end{table}

\subsection{Heat capacity}

\begin{figure}
\includegraphics[width=\figwp\columnwidth,viewport=0 150 582 819,clip]{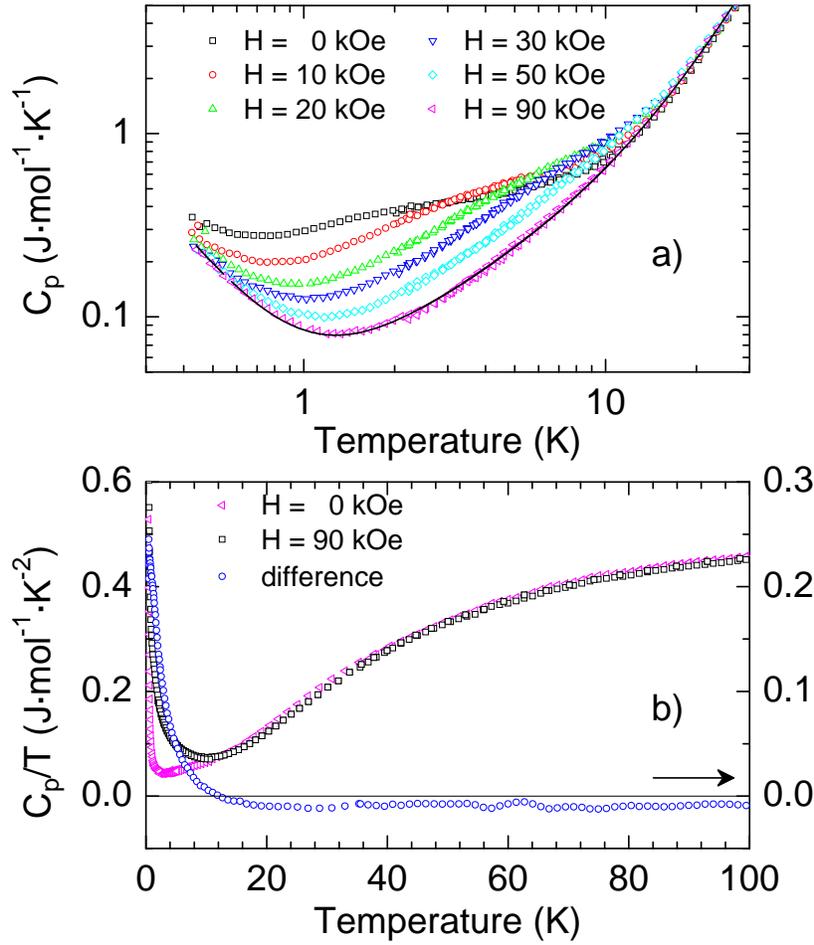}
\caption{Specific heat of \lacatbcodve\ under various magnetic fields. (a) The low-temperature $C_p$ in the
$log-log$ plot; the full line corresponds to analytical fit of the data at 90~kOe - $C_p=\alpha
T^{-2}+\beta T^3+\gamma T+\delta T^2$ with $\alpha=0.044$, $\beta=0.00019$, $\gamma=0.040$ and
$\delta=0.0006$. (b) $C_p$ divided by $T$ at zero field and 90~kOe over broader temperature range,
complemented with the difference graph in enlarged scale.} \label{FigCp}
\end{figure}

\begin{figure}
\includegraphics[width=\figwl\columnwidth,viewport=0 400 582 819,clip]{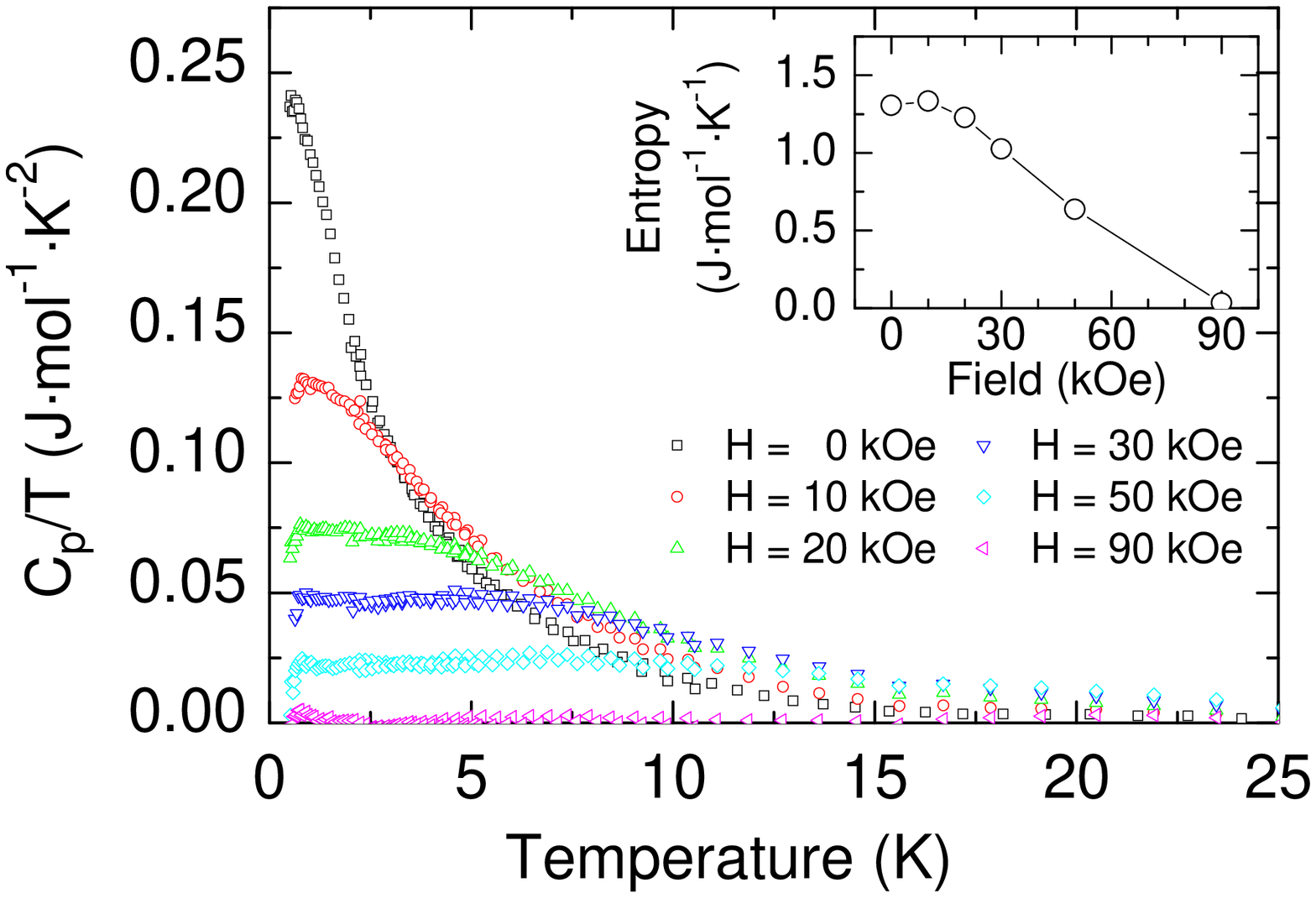}
\caption{The Tb$^{3+}$ related term of specific heat in \lacatbcodve. The inset
shows the calculated entropy change within the range $0.4-25$~K.} \label{FigCpTb}
\end{figure}

The heat capacity measurements on \lacatbcox\ did not show any observable $\lambda$-anomaly around
their Curie temperatures $T_C=80$~K, 55 K and 30 K for x = 0.1, 0.2 and 0.3, respectively. This
means that critical fluctuations associated with magnetic ordering are distributed over much
larger temperature range than it is for conventional ferromagnets. Let us mention an analogous
lack of criticality at apparent $T_C$ of Pr$_{0.7}$Ca$_{0.3}$CoO$_3$, which was attributed by
El-Khatib~\etal\ to a magnetic/electronic phase separation where FM clusters (presumably hole-rich
objects), preformed at temperatures as high as $\sim250$~K, coexisted with the hole-poor
paramagnetic matrix \cite{RefElKhatib2010PRB82_100411}.

The behaviour at the lowest temperatures is exemplified by the heat capacity data for
\lacatbcodve, presented in the upper panel of Fig.~\ref{FigCp}. There is a marked component that
is spanning below 10~K and seems to be proportional to the terbium doping. It decreases with
applied field and apparently vanishes in the maximum field of 90~kOe. The residual heat or
background can be fitted by a sum of common contributions, known for transition metal oxides. The
dominant term is the lattice heat $\beta T^3$, which is complemented with rather large linear
specific heat $\gamma T$ and marked Schottky anomaly at the lowest temperatures, $\alpha T^{-2}$.
To improve the fit, a small quadratic term $\delta T^2$ is included. Let us note that similar but
somewhat larger quadratic term was observed earlier for the phase separated systems \lasrcox\ with
x$<0.22$ and was attributed to the non-FM regions \cite{RefHe2009PRB80_214411}. Exact mechanism,
remained, however, unclear. In the same work the Schottky anomalous term $\alpha T^{-2}$ due to
$^{59}$Co nuclei experiencing hyperfine field in FM regions was also analyzed. It appears that the
present value of coefficient $\alpha=0.044$~J$\cdot$K$\cdot$mol$^{-1}$ is order of magnitude
larger, so that its nature should be different. We ascribe such enhanced $\alpha T^{-2}$ term to
the contribution of $^{159}$Tb nuclei present in FM regions, and argue by very large electronic
moment of Tb$^{3+}$ ions $\sim8.4\mu_B$ and hyperfine coupling constant 400~kOe/$\mu_B$, compared
to $\sim1.7\mu_B$ and 80~kOe/$\mu_B$ for Co$^{3+}$/Co$^{4+}$ in Ref.~\cite{RefHe2009PRB80_214411}.

Let us turn now to the specific heat excess that emerges in lower fields. In the difference graph
in the lower panel of Fig.~\ref{FigCp}, it is seen that the entropy associated with excessive heat
at zero field is seemingly transformed, at 90~kOe, to a larger linear specific heat that extends
at least up to 100~K, \textit{i.e.} well above Curie temperature of long-range ordered regions,
$T_C=55$~K. The actual change of coefficient $\gamma$ is about 0.010
J$\cdot$K$^{-2}$$\cdot$mol$^{-1}$, and we suppose it scales with applied field and modifies the
background. After subtraction the background terms, the excessive heat contribution is plotted in
$C_p/T$ $vs.$ $T$ graph in Fig.~\ref{FigCpTb}. The characteristic feature is the large roughly
constant part, which becomes obvious in applied fields and corresponds in fact to an additional
linear term $C_p$=$\gamma_m T$. The value of $\gamma_m$ decreases with increasing external field
whereas the temperature range of linear regime expands. We note that linear specific heat is a
general property of glassy systems, both the dielectrics and spin glasses. However, in the present
system this term should be attributed to Tb$^{3+}$ ions in FM regions, where they experience, even
in the absence of external field, a non-zero randomly oriented internal field. It is essential for
existence of such linear specific heat that the Tb$^{3+}$ ions possess finite moments of Ising
character \cite{RefCoey1978JPLETT39_L327}. This issue will be discussed below.

The entropy associated with Tb$^{3+}$ ions in \lacatbcodve\ can be calculated by integration of
excessive $C_p/T$. The data plotted in the inset of Fig.~\ref{FigCpTb} show that the entropy
change at zero field makes actually $\sim1.3$~J$\cdot$K$^{-1}$$\cdot$mol$^{-1}$. This value is in
reasonable agreement with theoretical $S_{total}=k\cdot \ln 2$ \textit{per} ion, which gives
1.15~J$\cdot$K$^{-1}$$\cdot$mol$^{-1}$ for population of 0.2~Tb$^{3+}$ \textit{per} f.u.
Nonetheless, the integral value over the temperature range of Fig.~\ref{FigCpTb} is not conserved
and decreases quickly with increasing external field. As mentioned above, the low-temperature
entropy is transferred to larger linear specific heat at intermediate temperatures or to a
modified stiffness of the system.

\section{Discussion}

Our comprehensive investigation of the \lacatbcodve\ system provides a significant amount of
information on the role of rare-earth ions. No signatures for eventual presence of Tb$^{4+}$ ions
are observed and we may also argue that this valence state is unfavourable since it would increase
much the size disorder on the perovskite A-sites. The rare-earth ions are thus in the Tb$^{3+}$
state in electronic configuration $4f^8$, which is a non-Kramers ion. The lowest lying free-ion
term is $^7F_{6}$ ($L=3$, $S=3$, $J=6$), and it is split by the low-symmetry crystal field to 13
singlets. The calculations available for TbAlO$_3$ of the same $Pbnm$ symmetry show that the
ground and first excited states of Tb$^{3+}$ ion are formed of 90\% by two conjugate
wavefunctions, $|6,+6\rangle+|6,-6\rangle$ and $|6,+6\rangle-|6,-6\rangle$, and their
eigenenergies differ by only 0.025~meV, representing a quasi-doublet
\cite{RefGruber2008JLUM128_1271}. This specific kind of accidental degeneracy has important
consequences. Firstly, a relatively modest magnetic field of external or molecular nature will mix
the eigenstates into a form of two pseudospins with the main weight of $|6,+6\rangle$ and
$|6,-6\rangle$, respectively. This results in large magnetic moment of Tb$^{3+}$ ions of about
8.4~$\mu_B$. Secondly, these moments have essentially an Ising-like character, which is a source
of large local anisotropy and makes \lacatbcodve\ the strongly coercitive system. This property is
reflected in the virgin magnetization curves by apparent "metamagnetic transition", and in the
hysteresis loop by remanent moment and coercive field both gradually increasing below $\sim20$~K.

The local Ising axes are oriented in the $ab$-plane of the $Pbnm$ perovskite structure. Since
there are two crystallographically equivalent sites of Tb$^{3+}$ ions related by reflection in the
$ac$-plane, their Ising axes make an angle $\pm \varphi$ with the orthorhombic axis $a$. The angle
$\varphi=36$\st\ has been determined for TbAlO$_3$ and, in the absence of spectroscopic data, we
suppose similar inclination also for present \lacatbcodve, though the energy splitting of
electronic levels might be here reduced due to smaller octahedral tilting. Let us note that the
$x$- and $y$-components of moments deduced from the neutron diffraction, $m_F=3.8\mu_B$/Tb and
$m_C=2.5\mu_B$/Tb, correspond in fact to $\varphi=34$\st, and a similar inclination is determined
also for \lacatbcox\ systems with other terbium dopings. What is varied are the magnitudes of the
ordered cobalt and terbium moments that rapidly decrease with $x$. The data in Table~\ref{TabMag}
show that Tb$^{3+}$ in $x=0.1$ exhibits the full theoretical moment $8.4\mu_B$/Tb, which signifies
a complete long-range ordering in this sample, while the moments observed by neutron diffraction
in samples with higher terbium contents, namely our target composition \lacatbcodve\ ($x=0.2$) and
sample $x=0.3$, drop gradually. By the comparison, we may conclude that the population of
long-range ordered domains ($>100$~nm) in the latter samples is reduced to 50\% and 33\%,
respectively.

The Ising character of Tb$^{3+}$ pseudospins elucidates also the character of low-temperature
specific heat of \lacatbcodve, namely it provides an explanation for the anomalously large linear
term. In the FM phase of mixed-valence cobaltites, each Tb$^{3+}$ ion experiences certain
molecular field causing Zeeman splitting of the quasi-doublet, $\Delta E=g\cdot\mu_B\cdot H_m$
where $g$ is the anisotropic gyromagnetic factor. The thermal population of the two electronic
levels gives a standard Schottky peak in specific heat. As pointed by Coey and von Molnar, when
molecular and applied fields are randomly oriented making an angle $\theta$ with respect to local
Ising axes, the Zeeman splitting varies as $\cos\theta$ and, instead of two sharp levels, the
system as a whole exhibits a continuous spectrum of excitations with constant density of states
from $\Delta E=0$ to $\Delta E_{max}$. Namely, the constant density of states is prerequisite for
the strictly linear specific heat \cite{RefCoey1978JPLETT39_L327}. We believe that the same
mechanism applies for present \lacatbcodve. Although the molecular field orientation at zero
external field will depend on details of the non-uniform FM phase, the randomness is guaranteed at
increased applied fields because of polycrystalline nature of our sample.

\begin{figure}
\includegraphics[width=\figwp\columnwidth,viewport=0 300 582 750,clip]{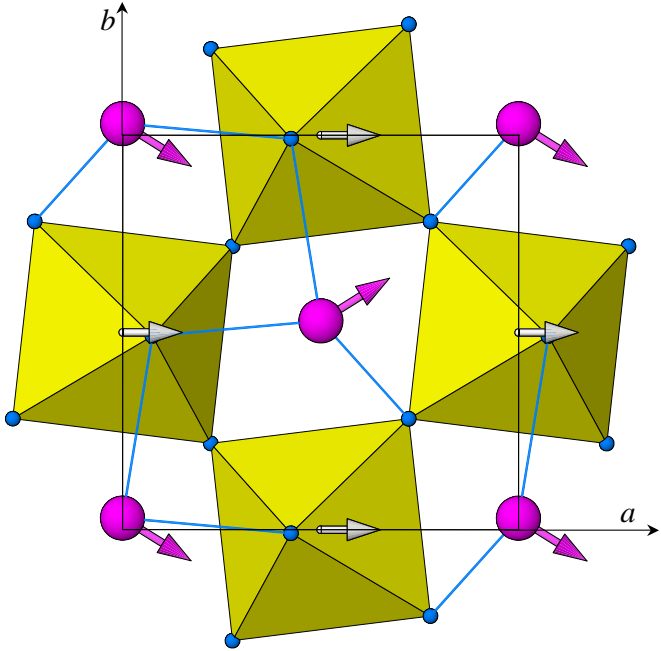}
\caption{Projection of \lacatbcodve\ structure along the $c$-axis, showing the F$_x$ ordering of
cobalt moments located in the $c=0$ level and the canted F$_x$C$_y$ ordering of rare-earth moments
located on the mirror plane at $c=0.25$. (The moment orientation in successive layers $c=0.5$ and
0.75 are identical.)}
 \label{FigStruktura}
\end{figure}

\section{Conclusions}

The perovskite cobaltite \lacatbcodve\ is a highly inhomogeneous chemical system with large size
disorder of A-site cations (La, Tb, Ca). The magnetic interactions of Co$^{3+}$/ Co$^{4+}$ ions
have clear prevalence of FM exchange, which is evidenced by FM-like susceptibility transition
though typical $\lambda$-anomaly in heat capacity is absent. We presume that the formation of FM
phase starts as nucleation of many FM regions that grow progressively and align according to
anisotropy axes defined by local strains and local fluctuations of charge carrier density. A
non-uniform magnetic state with coexistence of ordered and disordered regions, varying with the
temperature is thus anticipated. The neutron diffraction detects, nonetheless, a long-range FM
order of cobalt moments below $T_C=55$~K. This spontaneous alignment of cobalt spins is a source
of molecular field that acts on the Tb$^{3+}$ ions, which are randomly distributed over the A
sites of perovskite structure together with the non-magnetic La$^{3+}$ and Ca$^{2+}$ ions. As a
result of theis molecular field, the terbium pseudospins are gradually polarized and their static
arrangement becomes observable below $\sim20$~K. The experiments show that the effective
cobalt-terbium interaction is ferromagnetic, but the resulting arrangement of terbium moments is
canted due to alternating orientation of their local Ising axes. To summarize, the kind of
magnetic structure actually observed (see Fig.~\ref{FigStruktura}) and anomalous low-temperature
linear term of the specific heat point to a strong effect of Tb$^{3+}$ ions with large Ising-type
moments on the low-temperature magnetic behaviour of La,Ca-based cobaltites.

\ack This work was supported by Project No.~P204/11/0713 of the Grant Agency of the Czech
Republic. We acknowledge the Laboratoire Leon Brillouin (Saclay, France) for providing access to
the neutron beams and for all technical support during the experiments.

\section*{References}


\end{document}